\documentclass[trsc,nonblindrev]{informs3} 

\OneAndAHalfSpacedXI 

\usepackage[colorlinks=true,citecolor=black,linkcolor=blue]{hyperref}

\usepackage{etoolbox}
\makeatletter
\patchcmd{\maketitle}
 {\def\@makefnmark}
 {\def\@makefnmark{}\def\useless@macro}
 {}{}
\makeatother

\usepackage{subfig}

\usepackage{array,multirow,graphicx}

\usepackage{adjustbox}

\usepackage{natbib}
 \bibpunct[, ]{(}{)}{,}{a}{}{,}%
 \def\bibfont{\small}%
\TheoremsNumberedThrough     

\EquationsNumberedThrough    

\begin{document}


 \RUNAUTHOR{Papadopoulos et al.} 

\RUNTITLE{Personalized Pareto-Improving Pricing Schemes with Truthfulness Guarantees}

\TITLE{Integrated Traffic Simulation-Prediction System using Neural Networks with Application to the \\Los Angeles International Airport Road Network}

\ARTICLEAUTHORS{%
\AUTHOR{Yihang Zhang, Aristotelis-Angelos Papadopoulos\footnote{Corresponding author}, Pengfei Chen, Faisal Alasiri}
\vspace{-5pt}
\AUTHOR{Tianchen Yuan, Jin Zhou, Petros A. Ioannou}
\vspace{5pt}
\AFF{Ming Hsieh Department of Electrical and Computer Engineering, \\ University of Southern California, 
   Los Angeles, CA 90089 USA\\ \EMAIL{\{yihangzh, aristotp, pengfeic, alasiri, tianchey, jinzhou, ioannou\}@usc.edu} \URL{}}
} 

\ABSTRACT{%
Transportation networks are highly complex and the design of efficient traffic management systems is difficult due to lack of adequate measured data and accurate predictions of the traffic states. Traffic simulation models can capture the complex dynamics of transportation networks by using limited available traffic data and can help central traffic authorities in their decision-making, if appropriate input is fed into the simulator. In this paper, we design an integrated simulation-prediction system which estimates the Origin-Destination (OD) matrix of a road network using only flow rate information and predicts the behavior of the road network in different simulation scenarios. The proposed system includes an optimization-based OD matrix generation method, a Neural Network (NN) model trained to predict OD matrices via the pattern of traffic flow and a microscopic traffic simulator with a Dynamic Traffic Assignment (DTA) scheme to predict the behavior of the transportation system. We test the proposed system on the road network of the central terminal area (CTA) of the Los Angeles International Airport (LAX), which demonstrates that the integrated traffic simulation-prediction system can be used to simulate the effects of several real world scenarios such as lane closures, curbside parking and other changes. The model is an effective tool for learning the impact and possible benefits of changes in the network and for analyzing scenarios at a very low cost without disrupting the network.   
}%



\maketitle

%

\section{Introduction}
The complexity of the traffic network is immense due to the non-homogeneous dynamics of different vehicle classes at the vehicle level to traffic nonlinear behavior at the traffic flow level.  Mathematical models whether static \citep{patriksson}, dynamic \citep{dynamic_modeling} or stochastic \citep{8522045} used by most routing schemes cannot possibly capture the complexity of the real system in order to achieve the best possible outcomes. A true optimum route for a truck for example may end up being far away from the optimum generated from a model due to uncertainties not captured by the mathematical model that optimality is based on. The development of accurate mathematical models to describe traffic characteristics has always been a challenge. The availability of fast computers and advanced software tools allows for the first time the development of traffic simulation models which can provide the information and predicted states of the traffic network with much higher accuracy. With a mesoscopic traffic simulator DYNASMART \citep{jayakrishnan1994evaluation}, Mahmassani and Peeta developed a deterministic traffic assignment model with an iterative algorithm to gain a solution of system optimum objective with fixed demand \citep{mahmassani1993network,mahmassani1995system,peeta1995system}. In the work of \cite{ben1997development,ben1997simulation}, a traffic demand and supply simulator is used to generate user equilibrium route guidance for a traffic assignment system. The simulation-based method is also used to solve operational problems, such as the multimodal transport problem \citep{mahmassani2007dynamic,zhou2008dynamic,lu2009equivalent}. Recent successful applications of simulation-based methods on transportation systems are \citep{Abadi2016} and \citep{Zhao2018}, where they used a real-time traffic simulator as part of a centralized coordinated multimodal freight load balancing system and showed the significance of traffic simulators in planning freight routes to achieve a good balance of freight loads across the road and rail network. \cite{zhang2016combined} demonstrated the impact of the variable speed limit (VSL) control strategy on traffic flows using a microscopic traffic simulator built in PTV Vissim.\footnote{\href{https://www.ptvgroup.com/en/solutions/products/ptv-vissim/}{https://www.ptvgroup.com/en/solutions/products/ptv-vissim/}}  The simulator is able to show consistent results in terms of traffic flow levels and vehicle densities when compared to a macroscopic mathematical model.

In order to build a high quality traffic simulation network which can accurately capture the characteristics of the traffic network and flow under different traffic scenarios, it is crucial to estimate the Origin-Destination (OD) matrix, which measures the number of vehicles travel between different zones per unit time in the network. The OD matrix is an input to the traffic simulator and generates traffic flow on each link in the road network. Existing OD matrix estimation approaches can be classified into 2 different approaches. The first approach estimates the OD matrix by directly accessing the origin-destination information of individual vehicles through trip survey, GPS data of mobile devices and cellular positioning data of cell phone subscribers etc \citep{giaimo2002modifications, zhang2010daily, moreira2016time}. The other approach uses measured traffic flow data on the links of the road network to estimate the OD matrix indirectly \citep{yang1995heuristic,krishnakumari2020data,djukic2012efficient}. 

\cite{zhang2010daily} proposed an OD matrix estimation method by using the cellular positioning data. The cellular trajectories are obtained by recording all the signal-transition events and periodic location update events of cellular devices to determine the trip origins and destinations. \cite{moreira2016time} proposed a GPS-location data-based method to dynamically estimate the time-varying OD matrix of a traffic network by using a partitioning incremental framework. While the trip information-based methods are able to provide direct OD information, their applications are significantly limited by the cost of data and privacy concerns. 

\cite{yang1995heuristic} proposed a traffic flow measurement-based method which iteratively performs the traffic assignment and OD estimation in order to minimize the error in traffic flows. \cite{djukic2012efficient} first reduce the dimensionality of the OD matrix by performing Principal Component Analysis (PCA) and then estimate the reduced OD matrix using a colored Kalman filter. The traffic flow based approaches usually perform an inverse process of the traffic assignment operation using mathematical road network modal, which requires dynamic estimation of travel time of all routes in the network and iterative operation of the time-consuming traffic assignment.  

Since microscopic traffic simulations have the potential to adequately represent traffic realism and human behavior, the application of micro-simulation-based Dynamic Traffic Assignment (DTA) techniques has attracted considerable attention. \cite{barcelo2002heuristic} represented a heuristic approach to DTA, where two components were used to determine the path flow rates; one was based on a stochastic route choice method, and the other one was based on an approximation to dynamic user equilibrium conditions. The microscopic simulator (AIMSUN) was used for traffic network loading. The results of their case study showed a reasonably good agreement between the real system and the simulation model. A framework for a DTA model with an embedded microscopic simulation model was proposed in \cite{liu2005dynamic}. Two parts comprise the structure: the analytical dynamic user equilibrium (DUE) model and the micro-simulation software (Paramics). The method of successive averages was used to update path flows and ensure the convergence of the model. The performance of the proposed model was evaluated using a simulated network of the University of Arizona campus area at Tucson, Arizona. The results showed that for all the selected routes, the model achieved its equilibrium after several iterations. \cite{yang2017large} presented a highly detailed microscopic dynamic traffic assignment framework where the travel times of a traffic network were solved iteratively; the iterations continue in search of a stable Wardrop equilibrium. The traffic simulation software (TransModeler) for the network loading in each iteration was used to fully capture the dynamics and fidelity of real-world phenomena on both the demand and supply sides. Four real-world projects were presented to demonstrate the advantages of the microscopic DTA in practice.  

In this paper, we develop an integrated simulation-prediction system for the Los Angeles International Airport (LAX) based on real world traffic data. To calibrate the model based on the available flow data, we propose an OD matrix estimation scheme that uses a microscopic traffic simulator with a DTA scheme, an OD matrix generation procedure with feasibility guarantees based on an optimization formulation and a Neural Network (NN) that learns to map traffic flow data into OD matrices.

\section{Methodology}
In this section, we discuss common difficulties in OD estimation via traffic flow data and the methodologies we use to build the integrated simulation-prediction transportation system to address them. The difficulties in OD estimation can be divided into two parts: to estimate OD with the knowledge of traffic flows and to determine and validate the associative relations between traffic flows and OD matrices.

\begin{figure}[!ht]
  \centering
  \includegraphics[width=0.6\textwidth]{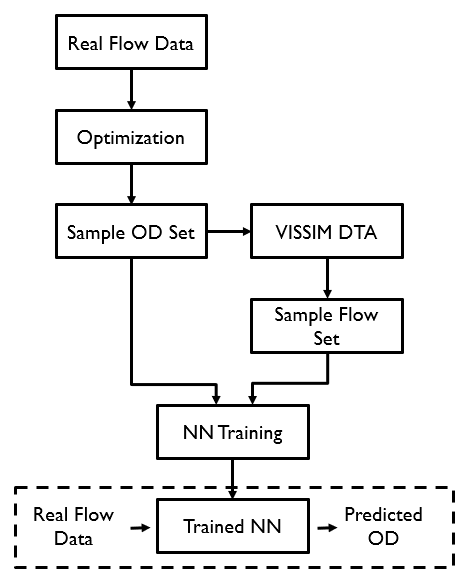}
  \caption{Structure of integrated simulation-prediction O/D estimation}\label{fig:structure}
\end{figure}

Estimating the OD matrix from flow measurement is difficult since there does not exist a one-to-one map from the flow rate to the OD matrix, that is, the same OD matrix can lead to different flow patterns. Therefore, we assume that the drivers in the traffic flow always behave in compliance with the dynamic traffic assignment (DTA) and the flow rates converge to an equilibrium very fast. Under this assumption, the map between the flow rates and the OD matrices becomes one-to-one, which makes is possible to estimate the OD matrix from the flow rate patterns. We implement a microscopic traffic model with PTV-Vissim to represent a simulator of the real road network. With the OD matrices as inputs, the simulator uses the DTA scheme to determine the routes for each OD pair, from which the aggregated traffic flows in each link can be gathered. Since the DTA assignment is based on reasonable behavior of the road network users, the routes from it are assumed to be close to the ones chosen by a driver. In short, given the knowledge of OD matrices, the simulator provides traffic flows close to the real ones. The simulator is also the base of validation. Unlike other estimation processes where the estimated variable is known a priori and can be compared directly, the OD estimate can only be compared based on generated traffic flows. In other words, to evaluate how good an OD estimate is, we need the function that maps an OD matrix to traffic flows. In our work, a traffic simulator is used for this purpose. Additionally, a Neural Network (NN) is used to estimate an OD matrix based on traffic flows. The overall OD matrix estimation method is shown in Figure~\ref{fig:structure}.

Real flow data are collected from sensors in the network. An optimization formulation with the flow conservation constraints is constructed based on the real flow data, whose solution is a set of OD matrices which are feasible given the real world flow data. We should note here that the OD matrices generated by solving the optimization problem are feasible only according to the flow data. The flow rate/OD matrix pairs are not guaranteed to follow the DTA assumption or any routing behavior. 

Then the OD matrices are fed into the traffic simulator, which adopts the Dynamic Traffic Assignment (DTA) scheme to predict the traffic flows in response to the input OD matrix. With the OD matrices achieved by solving the optimization problem and the corresponding traffic flows generated by the Vissim simulator using DTA, we train a Neural Network (NN) model with the traffic flows as input and the OD matrices as output. That is, the NN model is trained to mimic the inverse process of DTA, which will generate the DTA-compliant OD matrix given a set of flow measurement.

In the following sections, we describe in detail all the components used in the proposed OD estimation scheme shown in Figure~\ref{fig:structure}.   

\subsection{Optimization Formulation for Feasible OD Matrix Generation}
\label{subsec:od_generation}
\begin{figure}[!ht]
  \centering
  \includegraphics[width=0.6\textwidth]{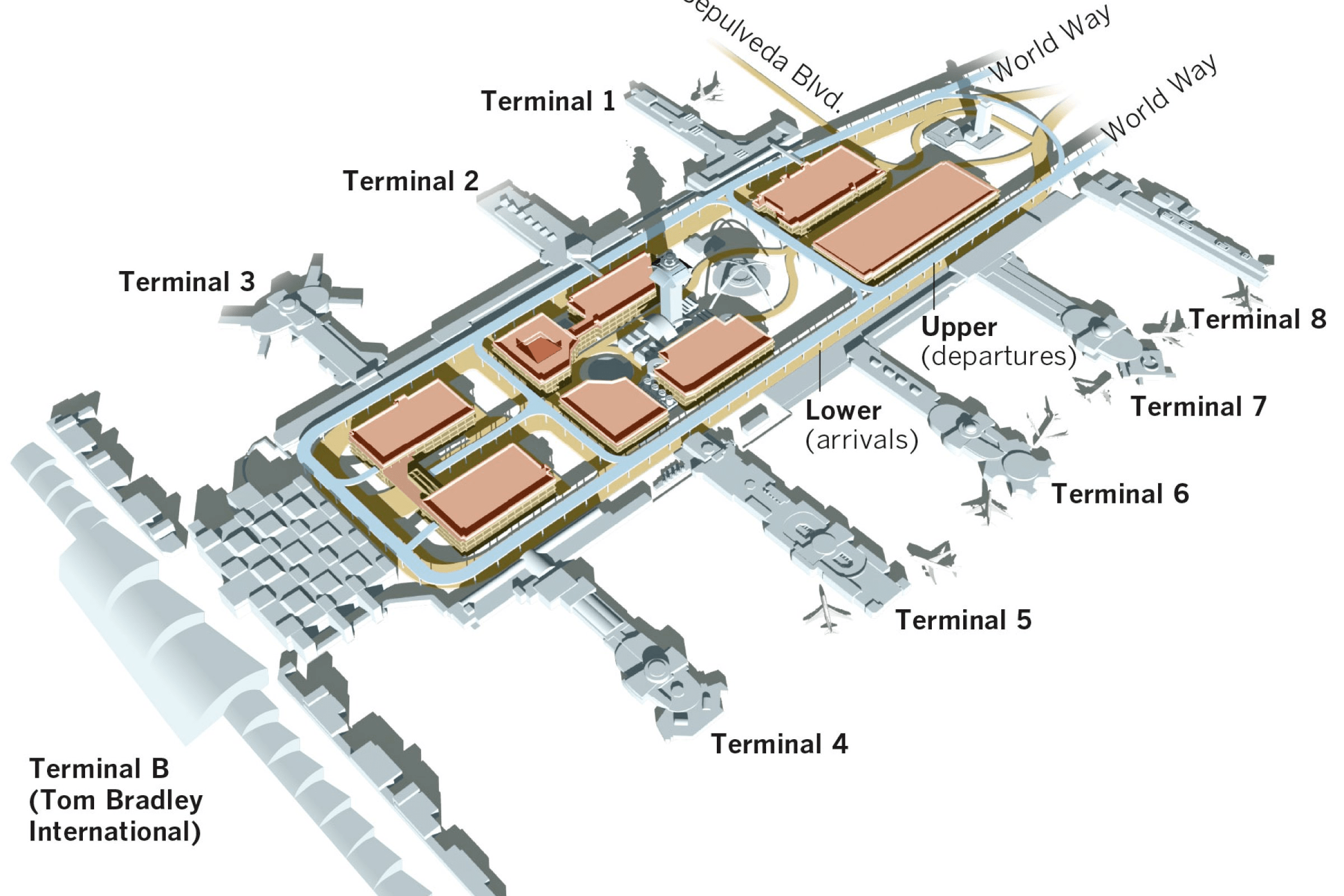}
  \caption{Central Terminal Area of the Los Angeles International Airport.\citep{LAXimeslos_angeles_times}}\label{fig:lax_3d_map}
\end{figure}

For the convenience of presenting the optimization formulation that generates feasible OD matrices, we first divide the Central Terminal Area (CTA) of the Los Angeles International Airport (LAX) into distinct zones. The CTA area of LAX consists of two levels as shown in Figure~\ref{fig:lax_3d_map}. Let $Z_1$ denote the upper level entrance, $Z_2$ denote the upper level exit, $Z_3$ denote the lower level entrance, $Z_4$ denote the lower level exit, $Z_5$ denote the upper level curb-side, $Z_6$ denote the lower level curb-side and finally $Z_7$ denote the rest of the Parking structures inside the LAX area.

Let $d_{ij}$ denote the demand from the origin $i$ to the destination $j$. An OD matrix $D$ consisting of elements $d_{ij}$ must satisfy the following constraints:

\begin{itemize}
    \item All demands are non-negative: \begin{equation}
    \label{first_constraint}
        d_{ij} \geq 0, \forall i,j
    \end{equation}
    
    \item All elements on the diagonal are equal to zero: \begin{equation}
    \label{second_constraint}
        d_{ii} = 0, \forall i
    \end{equation}
    
    \item The demand from the CTA exits is equal to zero: \begin{equation}
        d_{ij} = 0, \text{if} \hspace{5pt} i \in Z_2 \cup Z_4
    \end{equation}
    
    \item The demand to the CTA entrances is equal to zero: \begin{equation}
        d_{ij} = 0, \text{if} \hspace{5pt} j \in Z_2 \cup Z_4
    \end{equation}
    
    \item The demand from CTA entrances travelling directly to the CTA exits is equal to zero: \begin{equation}
        d_{ij} = 0, \text{if} \hspace{5pt} i \in Z_1 \cup Z_3\hspace{5pt} \text{and} \hspace{5pt}j \in Z_2 \cup Z_4 
    \end{equation}
    
    \item The demand between parking structures and curbside parkings is equal to zero: \begin{equation}
        d_{ij} = 0, \text{if} \hspace{5pt} i,j \in Z_5 \cup Z_6 \cup Z_7
    \end{equation}
    
    \item The demand from upper level (lower level) CTA entrances to lower level (upper level) curb parkings is equal to zero: \begin{equation}
        d_{ij} = 0, \text{if} \hspace{5pt} i \in Z_1 \hspace{5pt} \text{and} \hspace{5pt}j \in Z_6
    \end{equation}
    \begin{equation}
        d_{ij} = 0, \text{if} \hspace{5pt} i \in Z_3 \hspace{5pt} \text{and} \hspace{5pt}j \in Z_5
    \end{equation}
    
    \item The demand from upper level (lower level) curb parkings to lower level (upper level) CTA exits is equal to zero: \begin{equation}
        d_{ij} = 0, \text{if} \hspace{5pt} i \in Z_5 \hspace{5pt} \text{and} \hspace{5pt}j \in Z_4
    \end{equation}
    \begin{equation}
        d_{ij} = 0, \text{if} \hspace{5pt} i \in Z_6 \hspace{5pt} \text{and} \hspace{5pt}j \in Z_2
    \end{equation}
    
    \item The sum of demand at CTA entrances and CTA exits equals to flow rates at the entrances and exits: \begin{equation}
        \sum_j d_{ij} = q_i, \text{if} \hspace{5pt} i \in Z_1 \cup Z_3
    \end{equation}
    \begin{equation}
        \sum_i d_{ij} = q_j, \text{if} \hspace{5pt} j \in Z_2 \cup Z_4
    \end{equation}
where $q$ denotes the flow rate.
    
    \item For parking structures, the sum of flows of the upper level (lower level) entrances is equal to the sum of the demand from upper level (lower level) CTA entrances: \begin{equation}
        \sum_i d_{ij} = \sum q_{j,UL}, \text{if} \hspace{5pt} i \in Z_1 \hspace{5pt} \text{and} \hspace{5pt} j \in Z_7
    \end{equation}
    \begin{equation}
        \sum_i d_{ij} = \sum q_{j,LL}, \text{if} \hspace{5pt} i \in Z_3 \hspace{5pt} \text{and} \hspace{5pt} j \in Z_7
    \end{equation}
    
    \item For parking structures, the sum of flows of the upper level (lower level) exits is equal to the sum of the demand to upper level (lower level) CTA exits: \begin{equation}
        \sum_j d_{ij} = \sum q_{i,UL}, \text{if} \hspace{5pt} i \in Z_7 \hspace{5pt} \text{and} \hspace{5pt} j \in Z_2
    \end{equation}
    \begin{equation}
    \label{last_constraint}
        \sum_j d_{ij} = \sum q_{i,LL}, \text{if} \hspace{5pt} i \in Z_7 \hspace{5pt} \text{and} \hspace{5pt} j \in Z_4
    \end{equation}
    
\end{itemize}

The constraints (\ref{second_constraint})-(\ref{last_constraint}) are all linear and thus can be written in the form $Ad = b$, where $A$ is a constant matrix of coefficients, $b$ is a constant vector and $d$ is a vector consisting of variables $d_{ij}, \forall i,j$. To generate feasible OD matrices, we propose to solve the following optimization problem:  
\begin{equation}
\label{optim_for_OD}
\begin{aligned}
& \underset{d}{\text{minimize}}
& & ||Ad-b||^2 \\
& \text{subject to}
& & d_{ij} \geq 0, \; \forall i,j \end{aligned}
\end{equation}
where $||\cdot||$ represents the Euclidean norm. At this point, it is worth noting that we expect that the generated OD matrices are going to be sparse. The sparsity of the OD matrices is enforced through the feasibility constraints applied during the OD matrix generation process while solving the optimization problem (\ref{optim_for_OD}). The remaining zeros do not contain any useful information and therefore, to simplify the training process of the NN that we present later, we can vectorize
all the OD matrices into vectors that contain only nonzero entries. In the next subsection, we show how the generated OD matrices are used as an input to the traffic simulation model.

\subsection{Dynamic Traffic Assignment (DTA)}
As mentioned earlier, we can generate feasible OD matrices by solving the optimization problem (\ref{optim_for_OD}). These OD matrices serve as the input of the traffic simulation model. The output of the simulation model is the flow data of the designated links. According to the sensor distribution provided by LAX, these links are equipped with sensors that monitor the traffic in real-time. 

The simulation is carried out in microscopic level using PTV Vissim 10. Due to the large size of the LAX road network, Static Traffic Assignment (STA) with manually creating each route is tedious and unrealistic. Therefore we adopt the DTA approach to determine the route for each OD pair. In Vissim, the route decision depends on three factors, the travel time, the spatial length and the financial cost. For each link, a general cost can be computed as a weighted sum:
\begin{equation}
    C = \alpha T_{tr} + \beta L + \gamma C_f
\end{equation}
where $C$ is the general cost, $T_{tr}$ is the travel time, $L$ is the length of the link, $C_f$ is the financial cost and $\alpha, \beta, \gamma$ are weight parameters. The general cost of each route is simply the total cost of links from which the route is composed. There is only one best route for each OD pair. However, we assume that not all drivers choose the best route with the minimum cost but instead their route decisions follow a certain distribution among all known routes. In Vissim, the route distribution is based on the following logit function:
\begin{equation}
    p(R_j)=\frac{e^{-\eta \cdot log(C_j)}}{\sum_{i} e^{-\eta \cdot log(C_i)}}
\end{equation}
where $p(R_j)$ is the probability that route $j$ being selected, $\eta$ is the sensitivity parameter and $C_j$ is the general cost of route $j$.

\begin{figure}[!ht]
  \centering
  \includegraphics[width=0.7\textwidth]{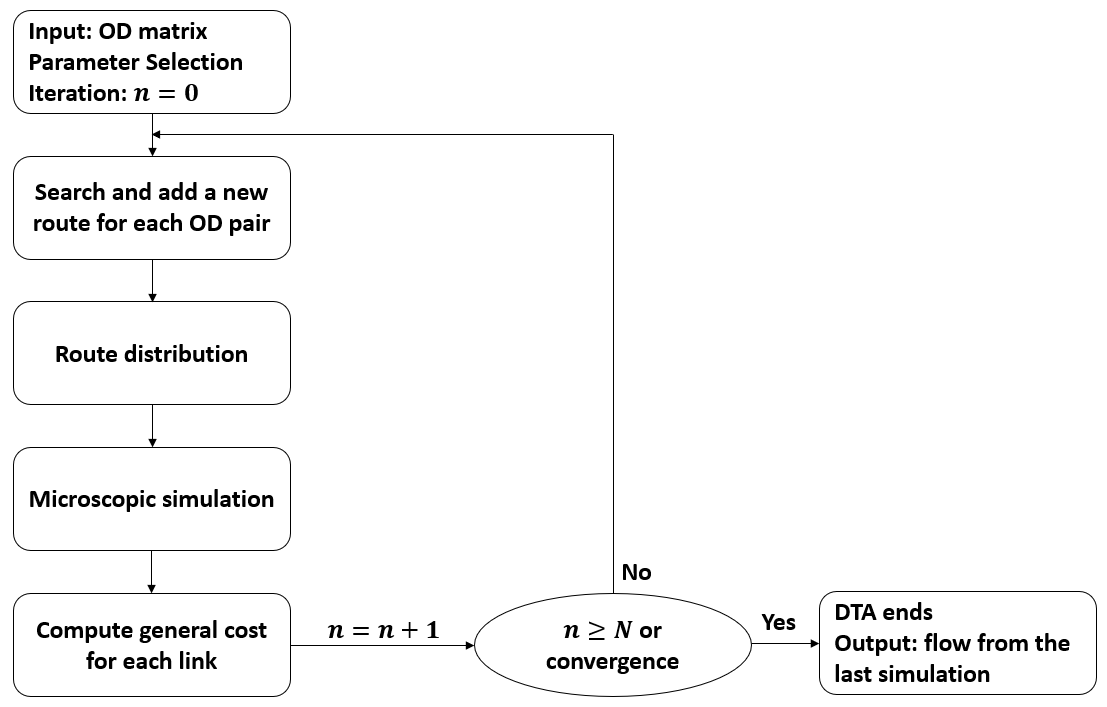}
  \caption{Flow Diagram of Dynamic Traffic Assignment.}\label{fig:DTA_flowchart}
\end{figure}

In order to complete the DTA for one OD matrix, we need to run the simulation multiple times until we reach the maximum iteration number $N$ or the convergence criterion is met. During each iteration, a new route with minimum cost is added to the route set for each OD pair. Then the demand is split onto the existing routes based on the logit function. The travel time and the financial cost of each link are updated after each simulation. Since there is no recorded travel time and financial cost in the first iteration, we can initialize the values as proportional to the length of the link. The convergence is checked at the end of each simulation and it also depends on the travel time and the financial cost of each link. We consider the convergence criterion is met when there is no significant change in these two quantities between two consecutive simulations. 
The diagram shown in Figure~\ref{fig:DTA_flowchart} describes the general procedure of DTA.

In the next subsection, we propose to train a Neural Network that uses the traffic flows generated by DTA as input training data and the corresponding OD matrices as output training data.

\subsection{Neural Network (NN)}
The training of the neural network model needs a loss function which quantifies the difference between the result we want and the current estimation result. In our work, we use the mean-squared error as the loss function. In order to avoid high weights, we add $l_1$ regularization term in the objective function \citep{10.5555/2380985} to penalize high weights. We also use cross-validation to select
hyperparameters such as the number of hidden layers, the dimension of each layer and the types of activation functions \citep{10.5555/1643031.1643047}. Dropout is added after the first dense layer to prevent overfitting \citep{10.5555/3086952}.
The feed-forward operation of the $l$-th layer can be written as:
\begin{equation}
    \label{feedforward_operation}
    z_{i}^{l+1} = {\bf \it W_{i}}^{l+1} {\bf y}^{l} + b_{i}^{l+1}
\end{equation}
\begin{equation}
    y_{i}^{l+1} = max(0,z_{i}^{l+1})
\end{equation}
where $W_{i}^{l+1}$ is the $i$-th row of the weight matrix at the $l$-th layer, $b_{i}^{l+1}$ is the $i$-th element of the bias vector at the $l$-th layer and $max(0,z_{i}^{l+1})$ is the RELU activation function. In the case where dropout is used, ${\bf y}^{l}$ is replaced by ${\bf \tilde y}^{l}$ in (\ref{feedforward_operation}) where ${\bf \tilde y}^{l} = {\bf y}^{l} \ast {\bf r}^{l}$ and $r_{j}^{l} \sim Bernoulli(p)$. The objective function that we minimize during training can be written as:
\begin{equation}
    L(\theta) + \lambda \Phi(\theta) = \frac{1}{M} \sum_{i=1}^{M} (y^{(i)} - \tilde y^{(i)})^2 + \lambda \Phi(\theta)
\end{equation}
where $\theta$ is the set of all parameters, $\Phi$ is the regularization term, $M$ is the size of the training set, $\tilde y^{(i)}$ is the output of the last layer of the NN and $y^{(i)}$ is the observed value on the training data.

\section{Experimental Results}
\subsection{Data Collection and Pre-processing}
In order to train the deep neural network which predicts the OD matrix with flow rate information, real-world hourly traffic flow data have been collected from 35 sensors distributed around
the LAX area, including 14 inductive loop sensors that measure the flow rate at each entrance and exit of the CTA area, and 21 transaction counters that count the number of vehicles entering and exiting each parking lot inside the CTA area. The dataset covers the entire month of April 2016. 

\begin{figure}[!ht]
  \centering
  \includegraphics[width=0.7\textwidth]{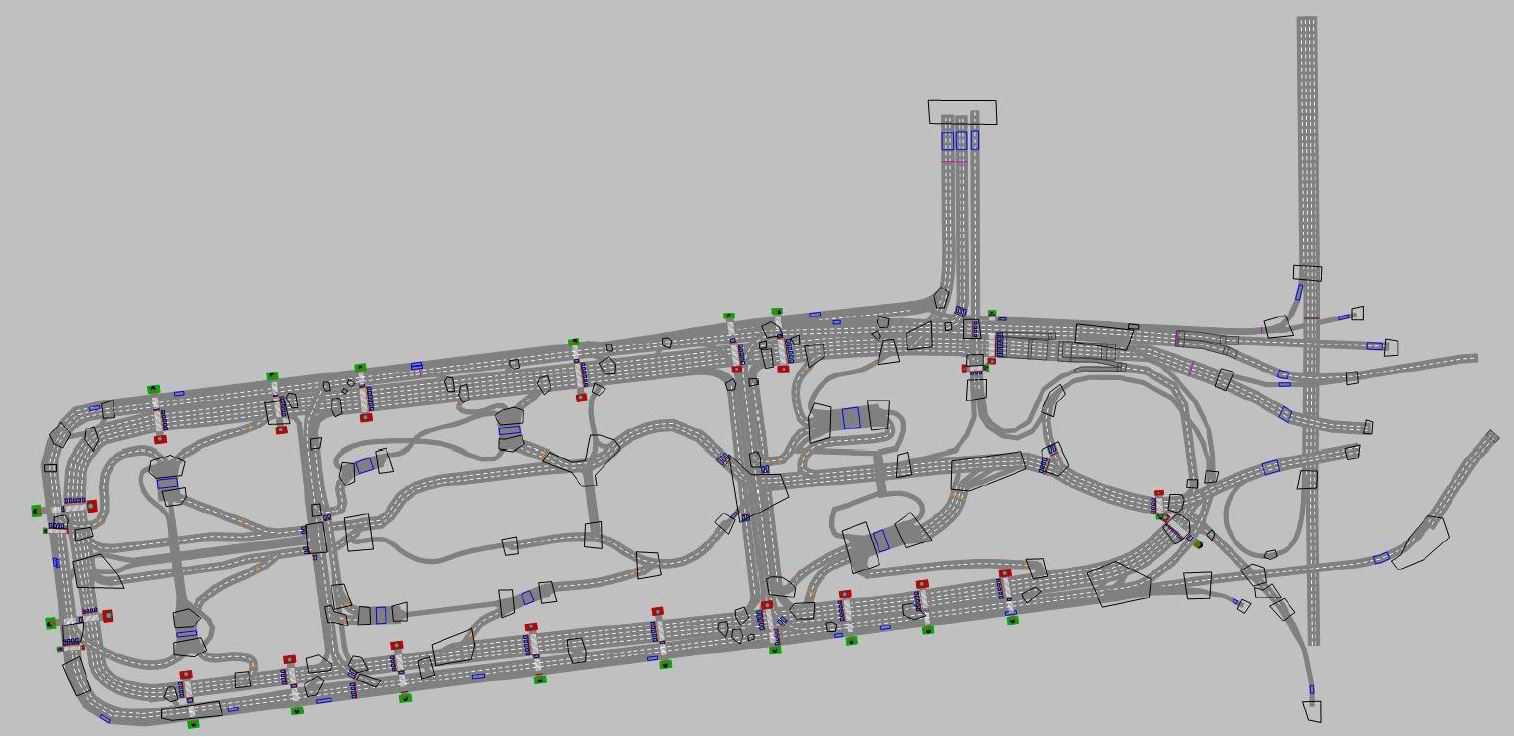}
  \caption{CTA area of LAX designed in PTV Vissim 10.}\label{fig:lax_at_vissim}
\end{figure}

In total, 32 OD zones are defined in this road network, including all entrance/exit groups, parking lots inside the CTA area and the curbside parking areas.  
Figure~\ref{fig:lax_at_vissim} from Vissim shows the details of this
network. The OD matrix records all the vehicle demands within the network in a certain
period of time, i.e. the $ij$-th entry representing the vehicle demand from zone i to zone
j. In our case the time period is one hour and
the OD matrix size is $32\times32$. As expected, the generated OD matrices are sparse, with only 161 nonzero entries out of 1024.

\subsection{NN Training}
Using the optimization problem (\ref{optim_for_OD}) for feasible OD matrix generation followed by the DTA assignment method performed in Vissim, we generated a total of 720 data samples. We used 600 samples for training and the rest 120 constituted the test set. We built a NN with one hidden layer consisting of 80 hidden units with a dropout rate 0.2. Both the hidden layer and the last layer used $l_1$ regularization with a value 0.02. Additionally, both layers used RELU. At this point, it is worth mentioning that the use of RELU in the last layer guarantees that we get values greater than zero at the output of the NN. This is essential in our design since the output of the NN expresses elements of the OD matrix which are always nonnegative.

During training, we minimize the Mean Squared Error (MSE) which is given by the following equation:
\begin{equation*}
    MSE_{NN} = \frac{1}{K}\sum_k \frac{1}{M}\sum_m (y_{km} - \tilde y_{km})^2
\end{equation*}
where $K$ is the number of the data samples, M is the dimensionality of the output vector of the neural network which in our case is 161, $y_{km}$ is the observed value of the $k$-th sample of the $m$-th element of the output vector and $\tilde y_{km}$ is the corresponding predicted value. The Adam optimization method \cite{DBLP:journals/corr/KingmaB14} was used during training with a learning rate 0.001. In total, we trained the NN for 50 epochs using a batch size of 96.

\begin{figure}[h]
\begin{center}
{\includegraphics[width=.45\textwidth]{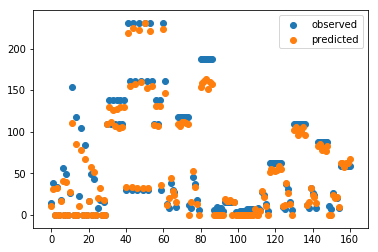}}
\includegraphics[width=.45\textwidth]{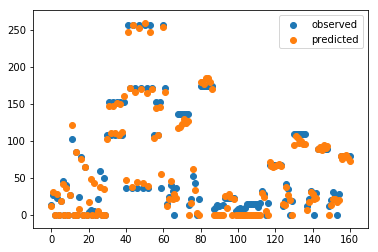}
\caption{Observed and predicted values for the 161 nonzero values of the OD matrix for: {\it Left:} Saturday, April 16th, 2016 at 2pm, {\it Right:} Monday, April 18th, 2016 at 10am.}
\label{fig:dots}
\end{center}
\end{figure}

To measure the performance of the NN, we additionally use the relative Root Mean Squared Error (rRMSE) which is given by the following equation:
\begin{equation*}
    rRMSE_{NN} = \frac{\sqrt{\frac{1}{K}\sum_k \frac{1}{M}\sum_m (y_{km} - \tilde y_{km})^2}}{\frac{1}{K}\sum_k \frac{1}{M}\sum_m y_{km}} *100\%
\end{equation*}
After training, we measured the MSE, the RMSE and the rRMSE using the 120 test samples. The results are shown below:
\begin{equation*}
    MSE_{NN} = 125.66, \; \; \;
    RMSE_{NN} = 11.21, \; \; \;
    rRMSE_{NN} = 21.99\%
\end{equation*}
These results indicate that the NN learned to predict the nonzero values of the OD matrix in a very satisfactory level. To visualize the predictions made by the NN, in Figure~\ref{fig:dots}, we plot the observed and predicted values for the 161 nonzero values of the OD matrix for a fixed time slot of 1 hour for two different days. As can be seen from Figure~\ref{fig:dots}, the predicted values at the output of the NN are close to the observed values from the OD generated data, indicating again that the NN learned to predict an OD matrix at the output given traffic flow data at its input. In the next subsection, we measure the performance of the overall OD estimation scheme.

\subsection{Performance of the OD Estimation Scheme}
\begin{figure}[h]
\begin{center}
{\includegraphics[width=.45\textwidth]{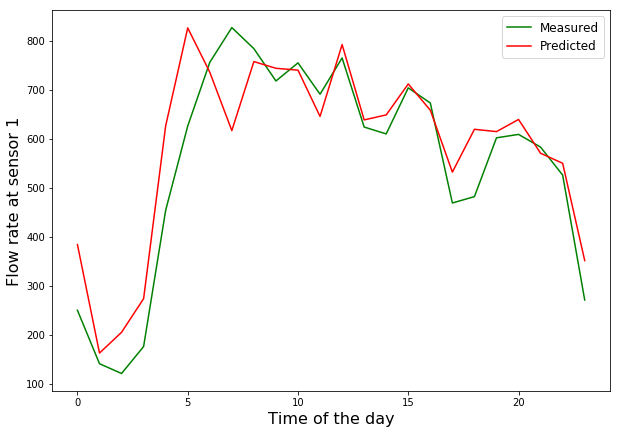}}
\includegraphics[width=.45\textwidth]{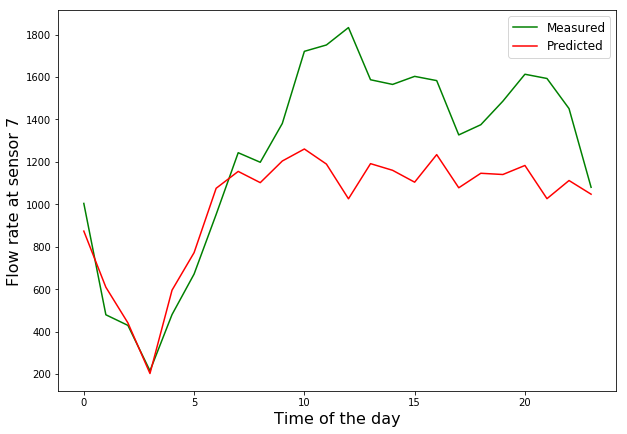}
\caption{Measured flow rates and flow rates predicted by the traffic simulator using the ODs estimated by the NN for different hours on Friday, April 1st, 2016 for: {\it Left:} Sensor location \#1, {\it Right:} Sensor location \#7.}
\label{predicted_flow}
\end{center}
\end{figure}

To measure the performance of the overall OD estimation scheme, we follow the following procedure. After the training process has been completed, we feed the real flow data into the NN model and obtain the estimated
OD vectors at the output. Note that we can easily rebuild OD matrices from the obtained OD vectors since the position of each nonzero demand is
known. To measure the goodness of these OD matrices, we feed them as input into the Vissim simulator and run DTA again. Then, we compare the flow data produced by running DTA in Vissim with the real flow data. To quantify the estimation error, we use the MSE, the RMSE and rRMSE which are defined below: 
\begin{equation*}
    MSE_{T} =  \frac{1}{S}\sum_{s=1}^{S} (q_{s} - \tilde q_{s})^2, \; \; \; \; RMSE_{T} = \sqrt{MSE_{T}}, \; \; \; \;
    rRMSE_{T} = \frac{RMSE_{T}}{\frac{1}{S}\sum_{s=1}^{S} q_{s}}*100\%
\end{equation*}
where $S$ is the total number of sensors which in our case is 35, $q_s$ is the real flow rate at sensor $s$ and $\tilde q_s$ is the flow rate at the output of Vissim after running the DTA using the estimated OD matrices produced by the NN. In Table~\ref{eval_OD_scheme}, we present the $MSE_T$, $RMSE_T$ and $rRMSE_T$ of the overall OD estimation scheme for different hours of a specific day. The day chosen is Friday, April 1st, 2016. Additionally, in Figure~\ref{predicted_flow}, we plot the real flow rates and the flow rates generated by Vissim using the ODs predicted by the NN for different hours of the same day and different sensor locations.

\begin{table*}[h]
\begin{center}
\begin{tabular}{|c|c|c|c|}
\hline 
\textbf{Hour}&\textbf{$MSE_{T}$}&\textbf{$RMSE_{T}$}&\textbf{$rRMSE_{T} (\%)$}\\
\hline
Midnight&2765.76&52.59&38.39\\
\hline
1am&1983.38&44.53&65.69\\
\hline
2am&877.90&29.63&50.93\\
\hline
3am&823.49&28.70&53.08\\
\hline
4am&6096.13&78.10&58.27\\
\hline
5am&7915.72&88.97&52.38\\
\hline
6am&22175.84&148.91&65.52\\
\hline
7am&14510.51&120.46&46.01\\
\hline
8am&13903.80&117.91&44.66\\
\hline
9am&23828.13&154.36&51.87\\
\hline
10am&31792.73&178.30&53.29\\
\hline
11am&33125.02&182.00&55.32\\
\hline
Noon&38346.02&195.82&62.66\\
\hline
1pm&20458.66&143.03&49.54\\
\hline
2pm&21368.97&146.18&49.91\\
\hline
3pm&28325.66&168.30&55.45\\
\hline
4pm&30117.82&173.54&60.32\\
\hline
5pm&7661.56&87.53&36.08\\
\hline
6pm&7888.82&88.82&34.21\\
\hline
7pm&19538.23&139.78&47.97\\
\hline
8pm&26651.24&163.25&50.69\\
\hline
9pm&37669.58&194.09&58.19\\
\hline
10pm&28973.38&170.22&61.98\\
\hline
11pm&4020.10&63.40&35.98\\
\hline
\end{tabular}
\end{center}
\caption{\label{eval_OD_scheme}$MSE$, $RMSE$ and $rRMSE$ of the overall OD estimation scheme for different hours on Friday, April 1st, 2016.}
\end{table*}

As can be observed from the results of Figure~\ref{predicted_flow}, the error between the real and the predicted flow rates varies according to different sensor locations or different times of the day. Furthermore, as can be seen from the results of Table~\ref{eval_OD_scheme}, there are times during the day where $rRMSE_T$ is around $35 \%$ and the overall OD estimation scheme performs well and times where $rRMSE_T$ is around $60 \%$ and the OD estimation scheme performs poorly. To further explore the performance of the proposed OD estimation scheme, we conduct an additional experiment. More specifically, we group the available sensors based on their median flow rate during the day as follows:
\begin{itemize}
    \item Low: Sensors with median flow rate [0, 36]
    
    \item Medium: Sensors with median flow rate (36, 175.5]
    
    \item High: Sensors with median flow rate (175.5, $max(flow rate)$]
\end{itemize}
Subsequently, we measure the $rRMSE_{T}$ for each group of sensors. The results are presented in Table~\ref{eval_OD_scheme_for_groups}. As can be seen from the results in Table~\ref{eval_OD_scheme_for_groups}, as the flow rate increases, the $rRMSE_{T}$ decreases demonstrating that the proposed OD evaluation scheme better estimates the flow rate of sensors whose median flow rate is medium or high. On the other hand, we observe that $rRMSE_{T}$ is high for sensors with low flow rate. Nevertheless, this does not significantly harm the performance of the overall OD estimation scheme since the absolute error remains low.

\begin{table*}[h]
\begin{center}
\begin{tabular}{|c|c|c|c|}
\hline 
\textbf{Hour}&\textbf{$rRMSE_{T,lo} (\%)$}&\textbf{$rRMSE_{T,me} (\%)$}&\textbf{$rRMSE_{T,hi} (\%)$}\\
\hline
Midnight&97.74&41.44&24.19\\
\hline
1am&207.96&133.33&36.16\\
\hline
2am&177.45&114.21&28.74\\
\hline
3am&59.92&96.37&31.87\\
\hline
4am&84.96&94.07&35.37\\
\hline
5am&97.75&96.90&31.06\\
\hline
6am&111.97&84.23&39.45\\
\hline
7am&90.59&49.05&28.05\\
\hline
8am&96.12&55.99&27.29\\
\hline
9am&95.18&59.98&32.29\\
\hline
10am&82.39&48.83&34.28\\
\hline
11am&93.17&38.70&35.62\\
\hline
Noon&110.84&48.06&39.74\\
\hline
1pm&82.03&43.43&31.70\\
\hline
2pm&89.22&40.96&31.63\\
\hline
3pm&86.06&47.32&35.19\\
\hline
4pm&93.20&54.85&38.05\\
\hline
5pm&89.24&30.85&22.68\\
\hline
6pm&65.44&38.96&21.46\\
\hline
7pm&87.16&42.62&30.55\\
\hline
8pm&82.72&51.76&32.50\\
\hline
9pm&89.65&53.58&37.73\\
\hline
10pm&106.56&50.51&40.88\\
\hline
11pm&100.97&26.10&23.42\\
\hline
{\bf Average}&{\bf 99.09}&{\bf 60.09}&{\bf 32.08}\\
\hline
\end{tabular}
\end{center}
\caption{\label{eval_OD_scheme_for_groups}$rRMSE$ of the overall OD estimation scheme for sensors with different traffic flow levels and for different hours on Friday, April 1st, 2016.}
\end{table*}

Even though the results demonstrate that the proposed OD evaluation scheme performs well, below, we discuss several factors that could improve its performance:

\begin{itemize}
    \item {\bf Limited training data.} Using the OD matrix generation procedure followed by the DTA method in Vissim, we generated a total of 720 input/output samples. Out of those, 600 were used for NN training and the rest were used for the performance evaluation of the NN. Generating more training data will improve the performance of the NN and will consequently improve the performance of the overall OD estimation scheme.
    \item {\bf Final objective.} The current formulation of the problem is two-stage with different objectives. Initially, we train a NN to predict OD vectors at the output by minimizing the Mean Squared Error (MSE) between the output OD vectors and the ODs generated by the optimization procedure. After training, we feed the real flow data to the NN, we get the predicted ODs at its output which we subsequently feed to Vissim that generates the corresponding flow data. In the final stage, we measure the performance of the OD estimation scheme using the MSE between the flow data generated by Vissim and the real flow data. In contrast, an end-to-end learning approach where we minimize the MSE between the flow data generated by Vissim and the real flow data and we update the weights of the NN based on the final objective will significantly improve the performance of the overall OD estimation scheme. However, this approach requires to analytically express the derivatives required to back-propagate the error from the flow generated by Vissim to the corresponding ODs at its input and therefore, we leave it as future work.   
    \item {\bf Temporal characteristics.} In our approach, a single-layer neural network was used to learn generating OD matrices at the output given traffic flow data at its input. Recurrent Neural Networks (RNNs) like the Long Short-Term Memory (LSTM) model \citep{hochreiter1997long} have been shown to be more efficient when dealing with sequence data since they can better capture the existing temporal characteristics. Therefore, we expect that the use of a LSTM model could increase the performance of the designed OD estimation scheme.
\end{itemize}

\subsection{One Lane Closed Scenario}
\begin{figure}[!ht]
  \centering
  \includegraphics[width=0.6\textwidth]{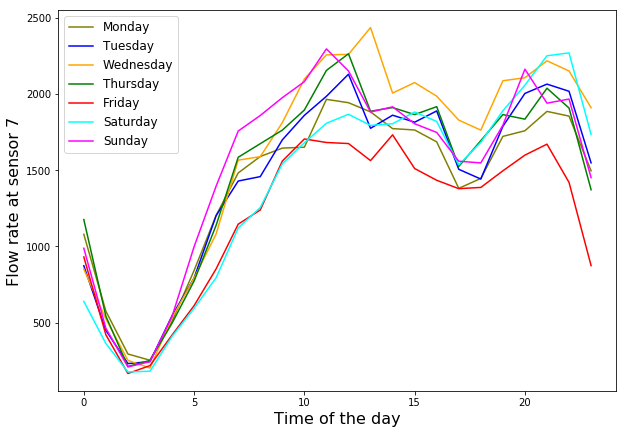}
  \caption{Flow rate per hour at sensor \#7 for the period Monday, April 11th until Sunday, April 17th, 2016.}\label{flow_rate_sensor_week}
\end{figure}

Los Angeles International Airport (LAX) is the world's third busiest and the United States' second busiest airport. According to statistics from \cite{LAWA}, LAX served more than 80 million passengers and more than 2 million tons of freight in 2016. In Figure~\ref{flow_rate_sensor_week}, we plot the flow rate per hour for a fixed sensor in the CTA area of the LAX airport for the period between Monday, April 11th and Sunday, April 17th, 2016. As can be observed from Figure~\ref{flow_rate_sensor_week}, the flow rate level in the CTA area of the LAX airport varies during different times in a day. However, during some time periods, the flow rate per hour is more than $2,000$. These numbers suggest that the LAX traffic agency has to take into account several factors that could affect traffic congestion levels inside the LAX airport and the areas nearby before making a decision. A microscopic traffic simulation model could capture the effect that different decisions can cause to the traffic congestion levels of the LAX airport.

\begin{figure}[!ht]
  \centering
  \includegraphics[width=0.7\textwidth]{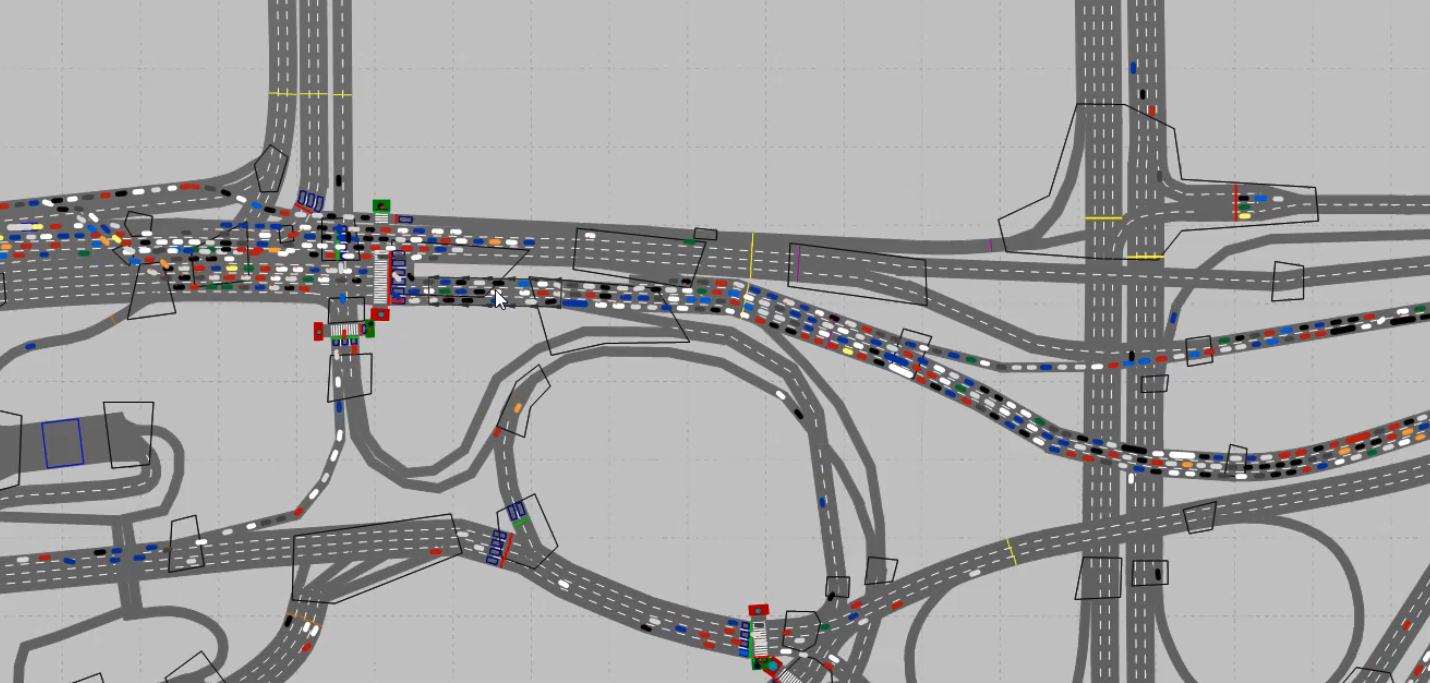}
  \caption{Lane closure simulated scenario at the entrance of the LAX airport.}\label{one_lane_closed_scenario}
\end{figure}

In this section, we simulate a scenario where there is a lane closure at an entry point of the LAX airport. As can be seen in Figure~\ref{one_lane_closed_scenario}, a lane closure at the entrance of the LAX airport during peak-hour traffic significantly increases the traffic congestion levels in the nearby area. Such a simulation model can be useful for the LAX traffic agency during their decision-making process since it visualizes the impact that a lane closure has on the LAX traffic.\footnote{We provide a video of the simulated lane closure scenario in this \href{https://drive.google.com/drive/folders/1RJZFeSlYCGIJcae2xgNS0B8ahFi4FgDn?usp=sharing}{link}.}   

\subsection{Curbside Parking Scenario}
\begin{figure}[!ht]
  \centering
  \includegraphics[width=0.7\textwidth]{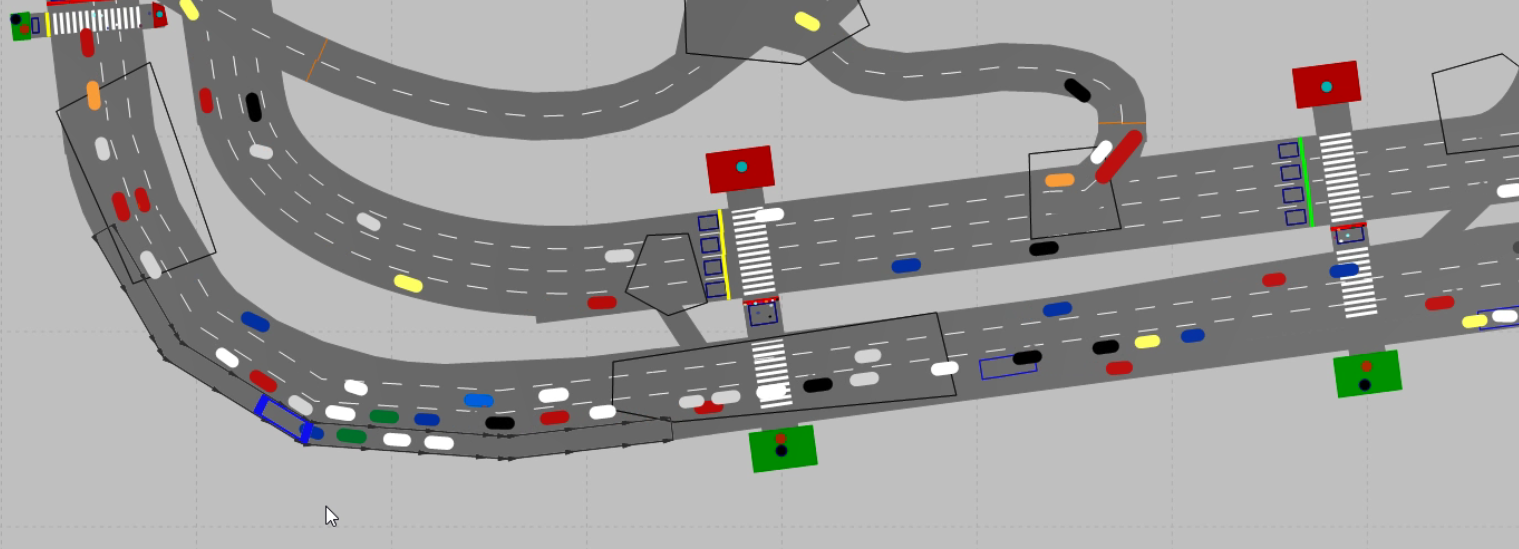}
  \caption{Curbside parking simulated scenario at Tom Bradley International Terminal of LAX airport.}\label{curb_side_parking_scenario}
\end{figure}

In this section, we simulate a scenario for curbside parking. In LAX airport, Transportation Network Companies (TNCs) such as Uber and Lyft are allowed to make a stop to pick-up or drop-off passengers only at designated curbside parking locations at the departures level (upper level). In Figure~\ref{curb_side_parking_scenario}, we present a curbside parking simulated scenario at Tom Bradley International Terminal of the LAX airport.\footnote{We provide a video of the simulated curbside parking scenario in this \href{https://drive.google.com/drive/folders/1RJZFeSlYCGIJcae2xgNS0B8ahFi4FgDn?usp=sharing}{link}.}

This scenario can be useful for the LAX traffic agency to make a decision as to whether it should allow Uber, Lyft and other TNCs to access the CTA area of the airport. In fact, at the end of 2019, the LAX traffic agency decided to ban Uber, Lyft and taxi pickups at the curb to alleviate the traffic congestion in the CTA area of the airport \citep{lax_bans_uber}.

\section{Conclusion}
In this paper, we developed an integrated simulation-prediction system  which estimates OD matrices and generates traffic flows that match those measured. The method is demonstrated for the Los Angeles International Airport (LAX) based on real world traffic data. To calibrate the simulation model based on the traffic flow data, we proposed an OD matrix estimation scheme that consists of three main components: an OD matrix generation method based on an optimization formulation, a Neural Network (NN) that learns to predict OD matrices at the output using traffic flow data as an input and a microscopic traffic simulator with a DTA scheme. Last, we demonstrated that the final integrated simulation-prediction system can be used to simulate real world traffic scenarios and can constitute an assistive tool for central traffic authorities in their planning and decision making process.  

\section{Acknowledgements}
This work has been supported by the Los Angeles World Airports (LAWA).  The authors would like to thank Mr Justin Erbacci and his team at LAWA for support and provided data. 

\section{Author Contributions}
The authors confirm contribution to the paper as follows: study conception and design: Y. Zhang, A.-A. Papadopoulos, P. Chen, F. Alasiri, T. Yuan, J. Zhou, P.A. Ioannou; data collection: Y. Zhang, A.-A. Papadopoulos, P. Chen, F. Alasiri, T. Yuan, J. Zhou; analysis and interpretation of results: Y. Zhang, A.-A. Papadopoulos, P. Chen, F. Alasiri, T. Yuan, J. Zhou, P.A. Ioannou; draft manuscript preparation: Y. Zhang, A.-A. Papadopoulos, P. Chen, F. Alasiri, T. Yuan, P.A. Ioannou. All authors reviewed the results and approved the final version of the manuscript.

\clearpage

%

%
%


\bibliographystyle{ifacconf.bst} 
\bibliography{ifacconf.bib} 

\begin{thebibliography}{32}
\providecommand{\natexlab}[1]{#1}
\providecommand{\url}[1]{\texttt{#1}}
\providecommand{\urlprefix}{URL }
\expandafter\ifx\csname urlstyle\endcsname\relax
  \providecommand{\doi}[1]{doi:\discretionary{}{}{}#1}\else
  \providecommand{\doi}{doi:\discretionary{}{}{}\begingroup
  \urlstyle{rm}\Url}\fi

\bibitem[{LAX(2019)}]{LAXimeslos_angeles_times}
 (2019).
\newblock Navigating lax and other southern california airports.
\newblock
  \url{https://www.latimes.com/travel/story/2019-11-19/navigate-lax-travel-holiday-airports-map}.
\newblock Accessed: 2020-07-23.

\bibitem[{Abadi et~al.(2016)Abadi, Ioannou, and Dessouky}]{Abadi2016}
Abadi, A., Ioannou, P.A., and Dessouky, M.M. (2016).
\newblock {Multimodal Dynamic Freight Load Balancing}.
\newblock 17(2), 356--366.

\bibitem[{Barcel{\'o} and Casas(2002)}]{barcelo2002heuristic}
Barcel{\'o}, J. and Casas, J. (2002).
\newblock Heuristic dynamic assignment based on microscopic traffic simulation.
\newblock In \emph{Proceedings of the 9th Meeting of the Euro Working Group on
  Transportation}, 1--16.

\bibitem[{Ben-Akiva et~al.(1997{\natexlab{a}})Ben-Akiva, Bierlaire, Bottom,
  Koutsopoulos, and Mishalani}]{ben1997development}
Ben-Akiva, M., Bierlaire, M., Bottom, J., Koutsopoulos, H., and Mishalani, R.
  (1997{\natexlab{a}}).
\newblock Development of a route guidance generation system for real-time
  application.
\newblock \emph{IFAC Proceedings Volumes}, 30(8), 405--410.

\bibitem[{Ben-Akiva et~al.(1997{\natexlab{b}})Ben-Akiva, Koutsopoulos,
  Mishalani, and Yang}]{ben1997simulation}
Ben-Akiva, M.E., Koutsopoulos, H.N., Mishalani, R.G., and Yang, Q.
  (1997{\natexlab{b}}).
\newblock Simulation laboratory for evaluating dynamic traffic management
  systems.
\newblock \emph{Journal of Transportation Engineering}, 123(4), 283--289.

\bibitem[{Chiu et~al.(2011)Chiu, Bottom, Mahut, Paz, Balakrishna, Waller, and
  Hicks}]{dynamic_modeling}
Chiu, Y.C., Bottom, J., Mahut, M., Paz, A., Balakrishna, R., Waller, T., and
  Hicks, J. (2011).
\newblock \emph{Dynamic Traffic Assignment: A Primer}.
\newblock Transportation Research Board.

\bibitem[{Djukic et~al.(2012)Djukic, Fl{\"o}tter{\"o}d, Van~Lint, and
  Hoogendoorn}]{djukic2012efficient}
Djukic, T., Fl{\"o}tter{\"o}d, G., Van~Lint, H., and Hoogendoorn, S. (2012).
\newblock Efficient real time od matrix estimation based on principal component
  analysis.
\newblock In \emph{2012 15th International IEEE Conference on Intelligent
  Transportation Systems}, 115--121. IEEE.

\bibitem[{Giaimo(2002)}]{giaimo2002modifications}
Giaimo, G.T. (2002).
\newblock Modifications to traditional external trip models.
\newblock \emph{Transportation research record}, 1817(1), 163--171.

\bibitem[{Goodfellow et~al.(2016)Goodfellow, Bengio, and
  Courville}]{10.5555/3086952}
Goodfellow, I., Bengio, Y., and Courville, A. (2016).
\newblock \emph{Deep Learning}.
\newblock The MIT Press.

\bibitem[{Hochreiter and Schmidhuber(1997)}]{hochreiter1997long}
Hochreiter, S. and Schmidhuber, J. (1997).
\newblock Long short-term memory.
\newblock \emph{Neural computation}, 9(8), 1735--1780.

\bibitem[{Jayakrishnan et~al.(1994)Jayakrishnan, Mahmassani, and
  Hu}]{jayakrishnan1994evaluation}
Jayakrishnan, R., Mahmassani, H.S., and Hu, T.Y. (1994).
\newblock An evaluation tool for advanced traffic information and management
  systems in urban networks.
\newblock \emph{Transportation Research Part C: Emerging Technologies}, 2(3),
  129--147.

\bibitem[{Kingma and Ba(2015)}]{DBLP:journals/corr/KingmaB14}
Kingma, D.P. and Ba, J. (2015).
\newblock Adam: {A} method for stochastic optimization.
\newblock In \emph{International Conference on Learning Representations}.
\newblock \urlprefix\url{http://arxiv.org/abs/1412.6980}.

\bibitem[{Kohavi(1995)}]{10.5555/1643031.1643047}
Kohavi, R. (1995).
\newblock A study of cross-validation and bootstrap for accuracy estimation and
  model selection.
\newblock In \emph{Proceedings of the 14th International Joint Conference on
  Artificial Intelligence - Volume 2}, 1137–1143.

\bibitem[{Krishnakumari et~al.(2020)Krishnakumari, van Lint, Djukic, and
  Cats}]{krishnakumari2020data}
Krishnakumari, P., van Lint, H., Djukic, T., and Cats, O. (2020).
\newblock A data driven method for od matrix estimation.
\newblock \emph{Transportation Research Part C: Emerging Technologies}, 113,
  38--56.

\bibitem[{LAWA(2017)}]{LAWA}
LAWA (2017).
\newblock Airport information - statistics.
\newblock
  \url{https://web.archive.org/web/20170211051258/http://lawa.org/welcome_lax.aspx?id=798}.
\newblock Accessed: 2020-07-23.

\bibitem[{Liu et~al.(2005)Liu, Ma, Ban, and Mirchandani}]{liu2005dynamic}
Liu, H.X., Ma, W., Ban, J.X., and Mirchandani, P. (2005).
\newblock Dynamic equilibrium assignment with microscopic traffic simulation.
\newblock In \emph{Proceedings. 2005 IEEE Intelligent Transportation Systems,
  2005.}, 676--681. IEEE.

\bibitem[{Lu et~al.(2009)Lu, Mahmassani, and Zhou}]{lu2009equivalent}
Lu, C.C., Mahmassani, H.S., and Zhou, X. (2009).
\newblock Equivalent gap function-based reformulation and solution algorithm
  for the dynamic user equilibrium problem.
\newblock \emph{Transportation Research Part B: Methodological}, 43(3),
  345--364.

\bibitem[{Mahmassani and Peeta(1993)}]{mahmassani1993network}
Mahmassani, H.S. and Peeta, S. (1993).
\newblock \emph{Network performance under system optimal and user equilibrium
  dynamic assignments: implications for ATIS}.
\newblock Transportation Research Board.

\bibitem[{Mahmassani and Peeta(1995)}]{mahmassani1995system}
Mahmassani, H.S. and Peeta, S. (1995).
\newblock System optimal dynamic assignment for electronic route guidance in a
  congested traffic network.
\newblock In \emph{Urban Traffic Networks}, 3--37. Springer.

\bibitem[{Mahmassani et~al.(2007)Mahmassani, Zhang, Dong, Lu, Arcot, and
  Miller-Hooks}]{mahmassani2007dynamic}
Mahmassani, H.S., Zhang, K., Dong, J., Lu, C.C., Arcot, V.C., and Miller-Hooks,
  E. (2007).
\newblock Dynamic network simulation--assignment platform for multiproduct
  intermodal freight transportation analysis.
\newblock \emph{Transportation Research Record}, 2032(1), 9--16.

\bibitem[{Moreira-Matias et~al.(2016)Moreira-Matias, Gama, Ferreira,
  Mendes-Moreira, and Damas}]{moreira2016time}
Moreira-Matias, L., Gama, J., Ferreira, M., Mendes-Moreira, J., and Damas, L.
  (2016).
\newblock Time-evolving od matrix estimation using high-speed gps data streams.
\newblock \emph{Expert systems with Applications}, 44, 275--288.

\bibitem[{Murphy(2012)}]{10.5555/2380985}
Murphy, K.P. (2012).
\newblock \emph{Machine Learning: A Probabilistic Perspective}.
\newblock The MIT Press.

\bibitem[{Nelson(2019)}]{lax_bans_uber}
Nelson, L.J. (2019).
\newblock Lax has banned uber, lyft and taxi pickups at the curb. here’s how
  the new system works.
\newblock
  \url{https://www.latimes.com/california/story/2019-10-28/uber-lyft-ban-lax-airport-pickup-terminal}.
\newblock Accessed: 2020-07-27.

\bibitem[{{Papadopoulos} et~al.(2019){Papadopoulos}, {Kordonis}, {Dessouky},
  and {Ioannou}}]{8522045}
{Papadopoulos}, A.A., {Kordonis}, I., {Dessouky}, M., and {Ioannou}, P. (2019).
\newblock Coordinated freight routing with individual incentives for
  participation.
\newblock \emph{IEEE Transactions on Intelligent Transportation Systems},
  20(9), 3397--3408.

\bibitem[{Patriksson(2015)}]{patriksson}
Patriksson, M. (2015).
\newblock \emph{The Traffic Assignment Problem - Models and Methods}.
\newblock Courier Dover Publications.

\bibitem[{Peeta and Mahmassani(1995)}]{peeta1995system}
Peeta, S. and Mahmassani, H.S. (1995).
\newblock System optimal and user equilibrium time-dependent traffic assignment
  in congested networks.
\newblock \emph{Annals of Operations Research}, 60(1), 81--113.

\bibitem[{Yang(1995)}]{yang1995heuristic}
Yang, H. (1995).
\newblock Heuristic algorithms for the bilevel origin-destination matrix
  estimation problem.
\newblock \emph{Transportation Research Part B: Methodological}, 29(4),
  231--242.

\bibitem[{Yang et~al.(2017)Yang, Balakrishna, Morgan, and
  Slavin}]{yang2017large}
Yang, Q., Balakrishna, R., Morgan, D., and Slavin, H. (2017).
\newblock Large-scale, high-fidelity dynamic traffic assignment: framework and
  real-world case studies.
\newblock \emph{Transportation research procedia}, 25, 1290--1299.

\bibitem[{Zhang et~al.(2010)Zhang, Qin, Dong, and Ran}]{zhang2010daily}
Zhang, Y., Qin, X., Dong, S., and Ran, B. (2010).
\newblock Daily od matrix estimation using cellular probe data.
\newblock In \emph{89th Annual Meeting Transportation Research Board},
  volume~9.

\bibitem[{Zhang and Ioannou(2016)}]{zhang2016combined}
Zhang, Y. and Ioannou, P.A. (2016).
\newblock Combined variable speed limit and lane change control for highway
  traffic.
\newblock \emph{IEEE Transactions on Intelligent Transportation Systems},
  18(7), 1812--1823.

\bibitem[{Zhao et~al.(2018)Zhao, Ioannou, and Dessouky}]{Zhao2018}
Zhao, Y., Ioannou, P.A., and Dessouky, M.M. (2018).
\newblock {Dynamic Multimodal Freight Routing Using a Co-Simulation
  Optimization Approach}.
\newblock \emph{IEEE Transactions on Intelligent Transportation Systems}.

\bibitem[{Zhou et~al.(2008)Zhou, Mahmassani, and Zhang}]{zhou2008dynamic}
Zhou, X., Mahmassani, H.S., and Zhang, K. (2008).
\newblock Dynamic micro-assignment modeling approach for integrated multimodal
  urban corridor management.
\newblock \emph{Transportation Research Part C: Emerging Technologies}, 16(2),
  167--186.

\end{thebibliography}


\end{document}